\documentclass[%
  reprint,
  amsmath,amssymb,superscriptaddress,
  aps,
  pra,
]{revtex4-2}
\usepackage{graphicx}
\usepackage{dcolumn}
\usepackage{bm}
\usepackage{color}
\usepackage{physics}
\usepackage{comment}
\usepackage{ulem}
\usepackage{color}
\usepackage[dvipsnames]{xcolor}
\usepackage{ulem, hyperref}
\usepackage{mathtools}
\usepackage{quantikz, upgreek}
\usepackage{tikz}
\usepackage{cancel}

\newcommand{\strike}[1]{%
  \ifmmode
    \cancel{#1}%
  \else
    \sout{#1}%
  \fi
}

\preprint{APS/123-QED}
\date{\today}
\begin{document}
  \title{Remote Gottesman-Kitaev-Preskill Entanglement Generation between Trapped Ions}
  \author{Haruki Aoki}
  \email{aoki.2.71828@keio.jp}
  \affiliation{School of Mathematical and Physical Sciences, Keio University, Yokohama, Kanagawa 223-8522, Japan}
  \author{Kazufumi Tanji}
  \affiliation{Department of Electronics and Electrical Engineering, Keio University, Yokohama, Kanagawa 223-8522, Japan}
  \author{Shohei Kiryu}
  \affiliation{Department of Electronics and Electrical Engineering, Keio University, Yokohama, Kanagawa 223-8522, Japan}
  \author{Masahiro Takeoka}
  \email{takeoka@elec.keio.ac.jp}
  \affiliation{Department of Electronics and Electrical Engineering, Keio University, Yokohama, Kanagawa 223-8522, Japan}
  \email{takeoka@elec.keio.ac.jp}
  \affiliation{Advanced ICT Research Institute, National Institute of Information and Communications Technology (NICT), Koganei, Tokyo 184-8795, Japan}
  \author{Junko Ishi-Hayase}
  \email{hayase@appi.keio.ac.jp}
  \affiliation{School of Mathematical and Physical Sciences, Keio University, Yokohama, Kanagawa 223-8522, Japan}
  \begin{abstract}
    %
    %
      Distributing entanglement directly between error-corrected logical qubits is an important capability for scalable quantum networks.
      We propose a heralded protocol for generating remote entanglement between Gottesman--Kitaev--Preskill qubits encoded in the motional modes of spatially separated trapped ions, using the internal electronic states to interface the motional qubits with emitted photons.
      We derive the protocol for single-rail and polarization photonic encodings and evaluate the effects of finite GKP squeezing, photon loss, photon recoil, motional heating, and motional dephasing on the decoded logical state.
      Moreover, we numerically analysed the fidelity of distributed logical Bell-pair fidelity.
      We find that one can achieve more than 95\% of fidelity in the realistic situation.
      The protocol combines motional-control and photonic-networking operations that have already been demonstrated in trapped-ion experiments.
  \end{abstract}
  \maketitle
  \section{Introduction}
    Entanglement between remote physical systems is a fundamental resource for quantum networks with quantum repeater~\cite{Wehner-Elkouss-Hanson2018, Awschalom-EtAl2021,Azuma-EtAl2023}, enabling applications such as distributed quantum computation~\cite{Sunami-EtAl2025a,Singh-EtAl2025a,Singh-EtAl2026}, and distributed quantum sensing~\cite{Liu-EtAl2021,Zhou-Brady-Zhuang2022,Kim-EtAl2024,Ueda-EtAl2026}.
    When two remote systems cannot interact directly, 
    entanglement swapping is a standard way to generate entanglement:
    each stationary qubit is first entangled with a flying photonic qubit, and a Bell-state measurement (BSM) on the photons heralds entanglement between the remote stationary qubits.
    Proof-of-principle experiments have been reported in various stationary qubit platforms, including trapped-ion systems~\cite{Casabone-EtAl2013, Hucul-EtAl2015, Krutyanskiy-EtAl2023a, Liu-EtAl2026c, Moehring-EtAl2007a, Nolleke-EtAl2013, OReilly-EtAl2024a, Saha-EtAl2025, Slodicka-EtAl2013, Stephenson-EtAl2020}, cold atoms~\cite{Chou-EtAl2005a, Chou-EtAl2007a, Hofmann-EtAl2012b, Li-EtAl2021, Liu-EtAl2024a, Ritter-EtAl2012a, vanLeent-EtAl2022, Yu-EtAl2020a}, and solid-state electron spin~\cite{Bernien-EtAl2013, Delteil-EtAl2016a, Hensen-EtAl2015a, Humphreys-EtAl2018a, Knaut-EtAl2024, Lago-Rivera-EtAl2021, Liu-EtAl2021a, Stolk-EtAl2024}.
    \par
    Despite these recent experimental progress, fidelity of the generated entanglement is still limited by qubit decoherence and by errors induced during photon generation.
    The dominant sources of stationary-qubit decoherence vary across physical platform; including magnetic field fluctuation in trapped ions~\cite{Bruzewicz-EtAl2019}, laser fluctuations of optical tweezers in neutral atoms~\cite{Henriet-EtAl2020}, and phonon or spin baths in solid-state spin systems~\cite{Doherty-EtAl2013}.
    In addition to such stationary-qubit decoherence, generation of qubit-photon entanglement may induce errors, for example, photon recoil~\cite{Apolin-Nadlinger2026, Kikura-EtAl2025, Saha-EtAl2025, Yu-EtAl2026}.
    Together, stationary-qubit decoherence and photon generation errors impose a practical limitation of fidelity on the generated remote entanglement.
    \par
    \par
    To overcome this limitation, the quantum error correction code (QECC) has been explored~\cite{Shor1995, Steane1996}.
    Alternatively, QECC can impove fidelity deterministically if physical operations and raw Bell state fidelity surpass certain threshold.
    Therefore, high fidelity Bell pair can be efficiently obtained by using QECC~\cite{Sunami-EtAl2026}.
    QECCs can be broadly classified into discrete-variable (DV) codes and bosonic codes which encode information in continuous-variable (CV) degree of freedom.
    DV-QECC schemes are deterministic and thus require less remote Bell pairs than distillation. 
    However, they still require a large number of physical qubits to encode a single logical qubit.
    
    \par
    In contrast to DV-QECC architectures, bosonic QECC, also known as QECC, encodes a logical qubit in a single oscillator mode with the infinite-dimensional Hilbert space.
    Consequently, it is possible to distribute a logical Bell pair with only two bosonic modes.
    \par
    The feasibility of using a single oscillator as a logical quantum memory is supported by rapid experimental progress in bosonic quantum control.
    In superconducting microwave cavities, experiments have advanced from the preparation of nonclassical states to logical-state preparation, active error correction, and universal oscillator control using cat~\cite{Cochrane-Milburn-Munro1999,Vlastakis-EtAl2013,Ofek-EtAl2016}, binomial~\cite{Michael-EtAl2016,Hu-EtAl2019}, and Gottesman--Kitaev--Preskill (GKP)~\cite{Gottesman-Kitaev-Preskill2001,Campagne-Ibarcq-EtAl2020,Eickbusch-EtAl2022} encodings.
    Trapped-ion motional modes have likewise progressed from GKP-state encoding to active error correction and universal logical control~\cite{Fluhmann-EtAl2019,deNeeve-EtAl2022,Matsos-EtAl2024,Matsos-EtAl2025}. 
    Atomic ensembles provide a complementary form of bosonic quantum memory, in which optical quantum states are mapped onto collective-spin excitations and subsequently retrieved~\cite{Honda-EtAl2008,Guo-EtAl2026}. These developments establish stationary bosonic modes as viable logical quantum memories.
    \par
    For remote entanglement distribution, however, such memories must also be interfaced with flying optical photons.
    Among the platforms considered here, atomic ensembles and trapped ions naturally combine a bosonic memory with optically addressable internal transitions. 
    These transitions can interface collective-spin and motional states, respectively, with flying optical photons, making both systems natural candidates for bosonic quantum-network nodes~\cite{Haussler-VanLoock2025}.
    \par
    \par
    Trapped-ion motional modes, already support key bosonic-code operations, as described above.
    The atomic ensembles, in contrast, an explicit protocol for generating entanglement between a stationary GKP qubit and a photonic qubit has yet to be developed. 
    \par
    Directly interfacing a motional state with a photonic state, however, would require strong coherent phonon–photon coupling~\cite{Parkins-Kimble1999}.
    Reaching this regime would require an improvement of approximately two orders of magnitude over current state-of-the-art ion–cavity QED systems~\cite{Takahashi-EtAl2020}. 
    Consequently, a practical route to remote motional GKP entanglement based on direct motion–photon coupling remains unavailable.
    \par
    In this paper, we propose a protocol for distributing GKP entanglement between remote trapped-ion motional modes mediated by the internal electronic states to interface the motional GKP qubits with the photonic qubits to circumvent the requirement for direct phonon--photon coupling.
    The protocol combines motional state control with photonic-networking operations on trapped ions, both of which have already been demonstrated experimentally.
    \par
    We further numerically analyse the effects of photon recoil, motional heating, and motional dephasing during the entanglement-generation process, and evaluate the decoded logical state by using stabilizer subsystem decomposition~\cite{Shaw-Doherty-Grimsmo2024a, Shaw-Doherty-Grimsmo2024}. 
    We report the resulting logical density matrices and their dependence on the noise parameters to characterize the proposed protocol.
    The result shows that the decoded logical Bell-state fidelity exceeds 95\% in experimentally relevent parameter regimes under current noise parameter.
    \par
    The remainder of this paper is organized as follows. 
    In Secs.~\ref{subsec:gkp_distribution}--\ref{subsec:evaluation}, we introduce the remote GKP Bell-state generation scheme, the noise model, and the evaluation method. 
    In Secs.~\ref{subsec:noiseless}--\ref{subsec:realistic-situation}, we analyze the performance of the proposed scheme under various conditions, including the noiseless case, photon loss, intrinsic noise in the trapped-ion system, and a realistic scenario incorporating all noise sources. 
    Finally, we conclude the paper in Sec.~\ref{sec:conclusion}.
  \section{Model and Methods}\label{sec:model-and-methods}
    \begin{figure}[t]
      \centering
      \includegraphics[width=\linewidth]{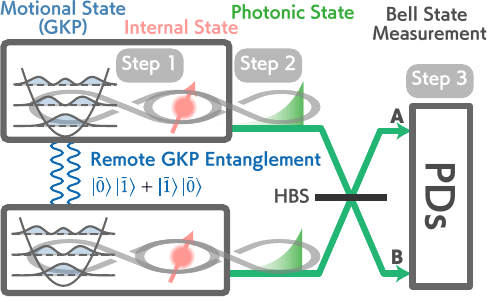}
      \caption{
          Overview of the heralded protocol for generating remote entanglement between motional GKP qubits in trapped ions.
          Each node contains a motional GKP qubit, an internal-state ancilla, and an optical interface to a flying photonic qubit.
          HBS denotes the half beam splitter to interfere the photons in Step~3.
          In Step~1, the motional GKP qubit is locally entangled with the internal state.
          In Step~2, state-selective excitation and photon emission generate a local motion--ion--photon GHZ state.
          In Step~3, the photons from the two nodes interfere at a central station and are measured by photo-detectors (PDs) to perform a Bell-state measurement.
          Conditioning on a successful photonic Bell-state measurement and subsequent local internal-state measurements generate the remote motional modes in a logical GKP Bell state.
      }\label{schematic}
    \end{figure}
    \subsection{GKP Entanglement Distribution} \label{subsec:gkp_distribution}
    Figure~\ref{schematic} outlines the proposed protocol for generating remote entanglement between motional GKP qubits.
    A trapped-ion system hosts two degrees of freedom: quantised motional modes in the trapping potential and internal electronic states of the ion.
    The motional mode stores the logical qubit encoded in a GKP code, in which all relevant operations, including state peparation, manipulation, and detection, are demonstrated by state-dependent forces.
      \par
\begin{figure}[t]
    \centering
    \includegraphics[width=\columnwidth]{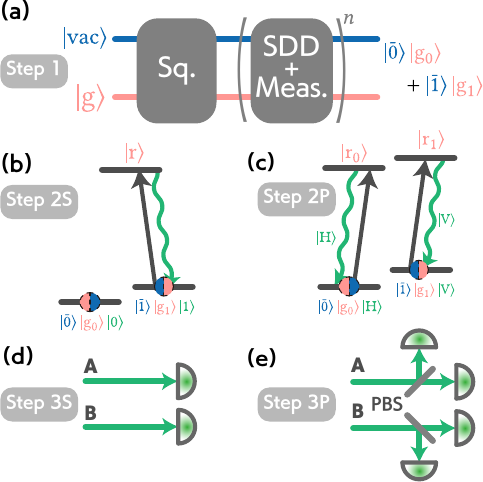}
    \caption{
        Detailed operations in Steps~1--3.
        (a) Conceptual representation of local motion--ion entangled state preparation from the motional vacuum and the internal ground state.
        Sq. and SDD + Meas. denote squeezing and state-dependent displacements and measurement implemented through motion--ion interactions.
        (b),(c) State-selective excitation for (b) the single-rail encoding and (c) the polarization encoding.
        (d),(e) Photonic Bell-state measurements for (d) the single-rail encoding and (e) the polarization encoding.
        The photodetectors are assumed to have insufficient temporal resolution to resolve the photon-arrival times.
    }\label{local-op}
    \end{figure}

    We now outline the three-step entanglement-distribution protocol as shown in Fig.~\ref{schematic}, which generates a remote Bell state between two motional GKP qubits.
    The local operations performed in each step are detailed in Fig.~\ref{local-op}.
    In Step 1, each node prepares a motional GKP logical state $\ket{\bar{0}}$ and entangles it with the internal state of the ion as $(\ket{\bar 0}\ket{\mathrm{g}_0}+\ket{\bar{1}}\ket{\mathrm{g}_1})/\sqrt{2}$, where $\ket{\bar{0}}, \ket{\bar{1}}$ denote motional state encoded by GKP code and $\ket{\mathrm{g}_0}, \ket{\mathrm{g}_1}$ denote internal state of ion.
    In Step 2, excitation and subsequent photon emission from the internal state convert this local motion--ion entanglement into a motion--ion--photon Greenberger-Horne-Zeilinger (GHZ) state.
    The motion--ion--photon GHZ state is necessary for remote entanglement generation because the motional GKP qubit does not directly couple to a photon. The internal state therefore mediates the interface between the motional qubit and the photonic qubit.
    In Step 3, the emitted photons from the two nodes are sent to a BSM node. 
    Conditioned on a successful BSM, the internal states are measured in a Pauli-X basis, disentangling the internal states and the two motional modes and resulting in the GKP Bell state.
      \par
    We further detail each step of generating the remote GKP Bell state below.
    In Step~1, the motional and internal degrees of freedom are coupled to the internal state of the ion through red- and blue-sideband interactions.
    In the interaction picture and setting $\hbar = 1$, the Hamiltonian is
    \begin{align}
      \hat{H}_{\mathrm{int}} = \Omega_{\mathrm{r}}\hat{a}^{\dag}_{\mathrm{m}}\hat{\sigma} + \Omega_{\mathrm{b}}\hat{a}^{\dag}_{\mathrm{m}}\hat{\sigma}^{\dag} + \mathrm{h.c.},\label{interaction-hamiltonian}
    \end{align}
    where, $\hat a_{\mathrm m}$ denotes the annihilation operator of the motional mode, and $\hat\sigma$ denotes the lowering operator of the internal two-level system.
    The parameters $\Omega_{\mathrm{r}}$ and $\Omega_{\mathrm{b}}$ are the coupling strengths of the red- and blue-sideband interactions, respectively.
    By choosing the amplitudes and phases of the red- and blue-sideband drives, one can realise state-dependent displacements of the motional mode, such as $\hat{D}(\beta\hat{\sigma}_x)$ and $\hat{D}(\mathrm{i}\gamma\hat{\sigma}_y)$.
    Using sequences of such bichromatic pulses, one can prepare motional GKP states and entangled states between the motional GKP qubit and the internal state following the methods of Refs.~\cite{Fluhmann-EtAl2019}. Figure~\ref{local-op} (a) shows the generation circuit of the GKP state and the ion-motion entanglement. 
    The motional and internal states are initially prepared in the vacuum state $\ket{\mathrm{vac}}$ and the ground state $\ket{\mathrm{g}}$, respectively . Since the approximated GKP state is represented by superposition of displaced squeezed states, the motional state is first squeezed through reservoir engineering~\cite{Kienzler-EtAl2015} and subsequently displaced by modular measurements~\cite{Fluhmann-EtAl2019}. As Pauli-X operation on the GKP state is defined by displacement, the entangling gate between the internal state and the motional GKP state is realised by state-dependent displacement. As an alternative way, the approximated GKP state can be generated by optimal pulse control~\cite{Matsos-EtAl2024}.
      \par
    In Step 2, the internal state is further entangled with an emitted photonic qubit.
    We consider two photonic encodings.
    The first is a single-rail encoding, in which the photonic qubit is encoded in the vacuum and single-photon states, $\ket{0}$ and $\ket{1}$. As shown in Fig.~\ref{local-op}(b), The single-rail photonic qubit is generated by exciting one of the ground states with a short $\pi$-pulse and the subsequent photon emission. The other ground state is isolated by applyng magnetic field generating Zeeman splitting.
    The second is a polarization encoding, in which the photonic qubit is encoded in the single-photon polarization states, $\ket{\mathrm{H}}$ and $\ket{\mathrm{V}}$. The polarization qubit is also generated by fast exicitation and spontaneous decay, but both ground states are involved as shown in Fig.~\ref{local-op}(c). By carefully choosing the transition, one can entangle the Zeeman qubit with the polarization qubit~\cite{Stephenson-EtAl2020}. 
    In both encodings, the emitted photon is defined by a temporal wavefunction with time width $\tau$,
    \begin{align}
      f(t) = \frac{1}{\sqrt{\tau}}\mathrm{e}^{-t/(2\tau)},
    \end{align}
    defined for $t\ge 0$.
    Since the single photon is generated by the short excitation and spontaneous emission, the temporal cost in Step 2 is limited by the temporal width $\tau$ of the emitted photon wavepacket, which is typically tens of nanoseconds~\cite{Saha-EtAl2025}.
    The temporal cost of the emission-based protocol is shorter than that of Raman-type photon-emission protocols, whose timescales are typically set by microsecond optical pulses~\cite{Kuhn-Hennrich-Rempe2002, Vasilev-Ljunggren-Kuhn2010}.
      \par
    After Step~2, the local motion--ion--photon entangled state for the single-rail encoding is
    \begin{align}
      \ket{\mathrm{GHZ}}_{\mathrm{S}} = \frac{1}{\sqrt{2}}\qty(\ket{\bar{0}}\ket{\mathrm{g}_0}\ket{0} + \ket{\bar{1}}\ket{\mathrm{g}_1}\ket{1:f}),\label{eq:ghz-single-rail}
    \end{align}
    where $\ket{1:f}$ denotes a single-photon state over a temporal mode $f(t)$ defined by
    \begin{equation}
        \ket{1:f}=\hat A_f^\dagger\ket 0=\int \dd t\ f(t)\hat a^\dagger(t)\ket{0},
    \end{equation}
    with the instantaneous creation operator $\hat a^\dagger(t)$.
    For the polarization encoding, the corresponding state is
    \begin{align}
      \ket{\mathrm{GHZ}}_{\mathrm{P}} = \frac{1}{\sqrt{2}}\qty(\ket{\bar{0}}\ket{\mathrm{g}_0}\ket{\mathrm{H}:f} + \ket{\bar{1}}\ket{\mathrm{g}_1}\ket{\mathrm{V}:f}).\label{eq:ghz-polarization}
    \end{align}
    Here, the polarization qubit is defined over the temporal mode $f(t)$,
    \begin{align}
        \ket{\Lambda:f}=\hat A_{f,\Lambda}^\dagger\ket 0=\int \dd t\ f(t)\hat a_{\Lambda}^\dagger(t)\ket{0},
    \end{align}
    for $\Lambda \in \{\text H,\text V\}$.
    A finite energy GKP state denoted by $\ket{\bar{\mu}}$ $(\mu\in\qty{0, 1})$ is defined as
    \begin{align}
      \ket{\bar{\mu}} \propto \sum_{n\in\mathbb Z}\exp\qty[-\frac{\pi}{2}\Delta^2(2n+\mu)^2]\hat{D}((2n+\mu)\alpha)\hat{S}(-\ln\Delta)\ket{\text{vac}} \label{gkp-def}
    \end{align}
    Here, $\hat D(\beta)=\exp(\beta\hat a_{\mathrm m}^{\dagger}-\beta^*\hat a_{\mathrm m})$, and we set $\alpha=\sqrt{\pi/2}$ in our quadrature convention.
    We parameterize the finite squeezing in decibels by $\Delta_{\mathrm{dB}}=-10\log_{10}(\Delta^{2})$.
    Equation~\eqref{gkp-def} represents a superposition of displaced squeezed states, where the Gaussian envelope $\mathrm{e}^{-2\pi n^2\Delta^2}$ suppresses components far from the origin in phase space, where $\Delta$ determines both the width of each squeezed peak and the envelope of the finite-energy GKP state.
    Since the GKP state stores quantum information as translation in a phase-space grid, small displacement errors can be detected and corrected by measuring the position and momentum.
    \par
    In Step~3, the photons from each node interfere at a half beam splitter (HBS).
    Each photonic mode is linearly transformed by
    \begin{align}
      \mqty(\hat{a}_{\mathrm{A}}'(t) \\ \hat{a}_{\mathrm{B}}'(t)) = \frac{1}{\sqrt{2}}\mqty(1 & 1 \\ 1 & -1)\mqty(\hat{a}_{\mathrm{A}}(t) \\ \hat{a}_{\mathrm{B}}(t)).
    \end{align}
    For the single-rail encoding, therefore, the state before photon-detection is 
    \begin{widetext}
        \begin{align}
            \ket{\psi}_{\mathrm{S}} &=\frac{1}{2}\ket{\bar{0}\mathrm{g}_0\bar{0}\mathrm{g}_0}_{\mathrm{AB}}\ket{00}_{\mathrm{AB}}+\frac{1}{2}\ket{\mathrm{GHZ}_4^{(+)}}_{\mathrm{AB}}\ket{1:f}_A\ket{0}_B + \frac{1}{2}\ket{\mathrm{GHZ}_4^{(-)}}_{\mathrm{AB}}\ket{0}_A\ket{1:f}_B\nonumber \\
            &+ \frac{1}{2\sqrt{2}}\ket{\bar{1}\mathrm{g}_1\bar{1}\mathrm{g}_1}_{\mathrm{AB}}(\ket{2:f}_{\mathrm{A}}\ket{0}_{\mathrm{B}}-\ket{0}_{\mathrm{A}}\ket{2:f}_{\mathrm{B}}),
        \end{align}
    \end{widetext}
    where the four-partite GHZ state $\ket{\text{GHZ}_4^{(\pm)}}_{AB}$ is defined by
    \begin{equation}
        \ket{\text{GHZ}_4^{(\pm)}}_{\mathrm{AB}}=\frac{1}{\sqrt 2}\left(\ket{\bar{1}\mathrm{g}_1\bar{0}\mathrm{g}_0}_{\mathrm{AB}} \pm \ket{\bar{0}\mathrm{g}_0\bar{1}\mathrm{g}_1}_{\mathrm{AB}}\right).
    \end{equation}
    For time-unresolved detection of a single photon,
    \begin{equation}
        \ketbra{1}=\int \dd t\ \hat{a}^\dagger (t)\ketbra{0} \hat{a}(t),
    \end{equation}
    the reusltant state is therefore $\ket{\text{GHZ}_4^{(\pm)}}$ depending on the detection pattern. By measuring the internal qubits in Pauli-X basis and corresponding feedforward operations, we obtain the GKP Bell state,
    \begin{equation}
        \ket{\Psi^{(+)}}_{\mathrm{AB}}= \frac{1}{\sqrt 2}\qty(\ket{\bar{0}\bar{1}}_{\mathrm{AB}} + \ket{\bar{1}\bar{0}}_{\mathrm{AB}}).
    \end{equation}
    For the polarization encoding,  the same logical Bell state is generated by assuming the identical temporal wavepackets, $f_{\mathrm{H}}(t) = f_{\mathrm{V}}(t) = f(t)$.
    \subsection{Noise Models}
    \begin{figure}[t]
    \centering
    \includegraphics[width=\columnwidth]{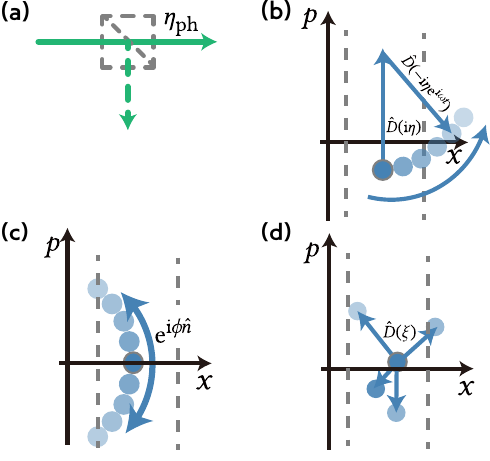}
    \caption{
        Noise channels included in the remote-entanglement model.
        (a) Photon loss is represented by a beam splitter with transmission efficiency $\eta_{\mathrm{ph}}$; it acts as amplitude damping for the single-rail encoding but reduces the heralding probability for the polarization encoding under two-photon postselection.
        (b) Photon recoil during Step~2 consists of the absorption displacement $\hat D(i\eta)$ followed by the emission displacement $\hat D[-i\eta e^{i\omega_{\mathrm m}t}]$.
        Because the photon-emission time is unresolved, the recoil operator $\hat R(t)=\hat D[-i\eta e^{i\omega_{\mathrm m}t}]\hat D(i\eta)$ is averaged over the emission-time distribution $f(t)$.
        (c) Motional heating during $t_{\mathrm{fly}}$ produces isotropic Gaussian random displacements $\hat D(\alpha)$.
        (d) Motional dephasing during $t_{\mathrm{fly}}$ produces random phase-space rotations $e^{i\phi\hat n}$ about the origin.
        In panels (b)--(d), the dark blue marker denotes an initial phase-space point and the faded markers denote possible outputs of the corresponding channel.
        The dotted boundaries indicate the GKP decision regions.
    }\label{fig:noise-channels}
    \end{figure}
    We analyse three imperfections related to motional GKP states: photon recoil, motional heating, and motional dephasing. We also consider the photon loss effect to compare the single-rail encoding with the dual-rail encoding as in Fig.~\ref{fig:noise-channels}(a).
    The phase-space schematics of photon recoil, motional heating, and motional dephasing are summarized in Fig.~\ref{fig:noise-channels}(b)--(d).
    For the single-rail encoding, photon loss acts as amplitude damping within the vacuum--one-photon qubit space.
    The photon loss channel with transmittivity $\eta_\text{ph}$ is modeled by
    \begin{align}
      \mathcal L_{\eta_{\mathrm{ph}}}(\rho) = \hat E_0\rho\hat E_0^\dagger
      +\hat E_1\rho\hat E_1^\dagger,
      \label{eq:photon-loss}
    \end{align}
    with
    \begin{equation}
        \begin{alignedat}{2}
          \hat{E}_0
          &= \ketbra{0} + \sqrt{\eta_{\mathrm{ph}}}\ketbra{1}
          &,\quad \hat{E}_1
          &=\sqrt{1-\eta_{\mathrm{ph}}}\ketbra{0}{1}.
        \end{alignedat}
    \end{equation}
    Here, $1-\eta_{\mathrm{ph}}$ is equivalent to the photon-loss probability.
    For the polarization encoding, both photonic states lie in the one-photon subspace.
    Loss therefore decreases the success probability but unchanges the fidelity of the generated entanglement.
    \par
    When two nodes are separated by a distance $L$, and a central station is placed midway between them, we define $t_{\mathrm{fly}}=L/(2v_{\mathrm{fly}})$, where $v_{\mathrm{fly}}$ is the photon propagation velocity in the channel.
    \par
    During Step 2, photon absorption and emission induce recoil errors for the motional GKP state~\cite{Kikura-EtAl2025}.
    When the driving pulse excites the ion from $\ket{g}$ to $\ket{r}$, the motional mode receives the photon momentum $\hbar k$ align with the motional mode axis.
    In phase space, this absorption recoil is represented by the displacement operator $\hat{D}(i\eta)$
    The Lamb--Dicke parameter $\eta=kx_0$ denotes the magnitude of the momentum kick on the selected motional mode, where $x_0$ is the position uncertainty along the recoil direction.
    When the excited state $\ket{\mathrm{r}}$ then spontaneously decays to $\ket{\mathrm{g}}$, the emitted photon kicks the motional state, represented by the displacement operator $\hat{D}(-i\eta e^{i\omega_{\mathrm{m}}t})$, where $\omega_{\mathrm{m}}$ is the motional trap frequency, and $t$ is the photon emission time.
    By combining the absorption with the emission recoil, We define the recoil operator
    \begin{align}
        \hat{R}(t) = \hat{D}\qty(-i\eta e^{i\omega_{\mathrm{m}}t})\hat{D}(i\eta).
    \end{align}

    Because the photon detection time is not resolved in our protocol, we average the density matrix over the temporal mode of the emitted photon and represent the recoil as an effective channel.
    For the single-rail encoding, only the case of single-photon emission undergoes the recoil.
    The corresponding recoil channel for one node is
    \begin{align}
    \label{eq:recoil-channel-sr}
        \mathcal{E}^{\mathrm{sr}}_{\mathrm{recoil}}(\rho) = \int_0^\infty \dd t_1\, |f(t_1)|^2 \hat{K}_{\mathrm{s}}(t_1)\rho\hat{K}_{\mathrm{s}}^{\dagger}(t_1),
    \end{align}
    with
    \begin{align}
        \hat{K}_{\mathrm{s}}(t_1) = \hat{I}_{\mathrm{m}}\otimes\ketbra{\mathrm{g}_0} + \hat{R}(t_1)\otimes\ketbra{\mathrm{g}_1}.
    \end{align}
    For the polarization encoding, both qubit states emit single-photons and thus undergo recoil, resulting in the two-photon recoil channel for one node,
    \begin{align}
        \label{eq:recoil-channel-dr}
        \mathcal{E}^{\mathrm{pol}}_{\mathrm{recoil}}(\rho) = \int_0^\infty \dd{t_0} \int_0^\infty \dd{t_1}\abs{f(t_0)}^2\abs{f(t_1)}^2 \notag \\ 
        \times \hat{K}_{\mathrm{d}}(t_0, t_1)\rho \hat{K}_{\mathrm{d}}^{\dagger}(t_0, t_1),
    \end{align}
    where
    \begin{align}
    \hat{K}_{\mathrm{d}}(t_0, t_1) = \hat{R}(t_0)\otimes\ketbra{\mathrm{g}_0} + \hat{R}(t_1)\otimes\ketbra{\mathrm{g}_1}.
    \end{align}
    Here, we assumed the Lamb--Dicke parameter is independent of polarization.
    
    \par
    In addition to photon-recoil errors, the motional GKP states are continuously decohered by motional heating and dephasing until measuring them. 
    For long distance communication, the propagation time $t_\text{fly}$ is dominant in the whole process. Therefore, we approximately characterize these two decoherence by $t_\text{fly}$.
    This heating is commonly attributed to electric-field noise at the ion position, originating from technical noise and surface noise of the trap electrodes.
    Motional heating is defined as a Gaussian random-displacement channel,
    \begin{align}
    \mathcal{E}_{\mathrm{heat}}(\rho) = \frac{1}{\pi\dot{\bar{n}}t_{\mathrm{fly}}}\int\dd[2]{\alpha}\mathrm{e}^{-\abs{\alpha}^2/(\dot{\bar{n}}t_{{\mathrm{fly}}})}\hat{D}(\alpha)\rho\hat{D}^{\dag}(\alpha),\label{eq:heating-channel}
    \end{align}
    where, $\dot{\bar n}$ is the motional heating rate and $\dot{\bar n}t_{\mathrm{fly}}$ is the mean occupation added during $t_\text{fly}$.
    \par
    Motional dephasing arises from fluctuations of the radio-frequency in trapping potential.
    This process is modelized by Gaussian phase diffusion,
    \begin{align}
    \label{eq:dephasing-channel}
    \mathcal{E}_{\mathrm{deph}}(\rho) = \frac{1}{\sqrt{2\pi\kappa_{\phi}t_{\mathrm{fly}}}}\int_{-\infty}^{\infty}\dd{\phi}\exp\qty(-\frac{\phi^2}{2\kappa_{\phi}t_{\mathrm{fly}}})\mathrm{e}^{i\phi\hat{n}}\rho\mathrm{e}^{-i\phi\hat{n}},
    \end{align}
    where $\kappa_\phi t_{\mathrm{fly}}=\operatorname{Var}(\phi)$ denotes phase variance accumulated during $t_\text{fly}$.
    \subsection{Evaluation}\label{subsec:evaluation}
    We use stabilizer subsystem decomposition (SSSD) to evaluate the remote GKP Bell state after ideal GKP decoding~\cite{Shaw-Doherty-Grimsmo2024a, Shaw-Doherty-Grimsmo2024}.
    SSSD provides a way to compute the logical state $\rho_{\mathrm{L}}$ obtained by ideal GKP decoding without explicitly performing the full decoding operation.
    For the bipartite GKP state, the ideally decoded two-qubit logical state can be reconstructed from expectation values of logical Pauli operators $\hat P_\text{L,i} \in \qty{\hat{I}, \hat{X}, \hat{Y}, \hat{Z}}$ as
    \begin{align}
    \rho_{\mathrm{L}} = \frac{1}{4}\sum_{i, j \in \qty{0, 1, 2, 3}} \ev{\hat{P}_{\mathrm{m}, i}\hat{P}_{\mathrm{m}, j}} \hat{P}_{\mathrm{L}, i}\otimes \hat{P}_{\mathrm{L}, j}. \label{logical-state-tomography}
    \end{align}
    Here, $\hat{P}_{\mathrm{m}, i}$ denotes the modular observable whose expectation value corresponds to the logical Pauli operator after ideal GKP decoding, and is defined by
    \begin{align}
      \hat{P}_{\mathrm{m}, i} = 
      \begin{dcases}
          \hat{1} & i = 0, \\ 
          \frac{1}{\pi}\sum_{n} \frac{(-1)^n}{n + 0.5}\hat{D}((2n + 1)\alpha) & i = 1,  \\ 
          \frac{1}{\pi^2}\sum_{n, m} \frac{\hat{D}((2n + 1)\alpha + i(2m + 1)\alpha)}{(n + 0.5)(m + 0.5)} & i = 2, \\ 
          \frac{1}{\pi}\sum_{n} \frac{(-1)^n}{n + 0.5}\hat{D}(i(2n + 1)\alpha) & i = 3.
      \end{dcases}
    \end{align}
    \par
    
    
    Since the logical Pauli operators are described by sum of displacement operator, we evaluate these expectation values using the Wigner characteristic function
    \begin{align}
    \chi(\gamma) = \tr\qty[\hat{D}(\gamma)\rho].\label{eq:characteristic-function}
    \end{align}
    \par
    For the Gaussian heating channel in Eq.~\eqref{eq:heating-channel}, the characteristic function transforms as
    \begin{align}
    \chi_{\mathrm{heat}}(\gamma) = \chi(\gamma)\mathrm{e}^{-\dot{\bar{n}}t_{\mathrm{fly}}\abs{\gamma}^2}.
    \end{align}
    Since the remained noises cannot be treated in the characteristic function, their effects are calculated numerically.
    \par
    To quantify the generated remote GKP entanglement, we compute the logical Bell-state fidelity following same manner as Refs.~\cite{Nielsen-Chuang2010,Lambert-EtAl2026},
    \begin{align}
    F_{\mathrm{L}} = \sqrt{\mel**{\Psi^{(+)}}{\rho}{\Psi^{(+)}}}.
    \end{align}
    \section{Results}
    \subsection{Baseline performance without physical noise}\label{subsec:noiseless}
    \begin{figure}[t]
    \centering
    \includegraphics[width=\columnwidth]{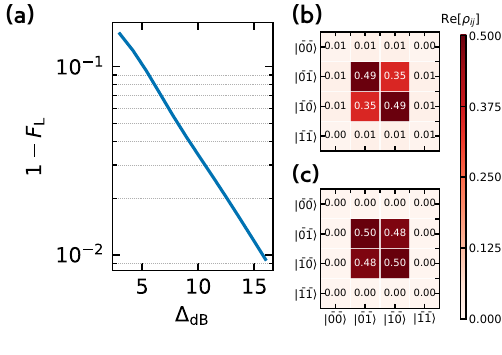}
    \caption{
      Baseline logical performance of the remote-entanglement protocol in the absence of physical noise.
      Panel (a) shows the logical Bell-state infidelity $1-F_{\mathrm L}$ as a function of the GKP squeezing level $\Delta_{\mathrm{dB}}$.
      Panels (b) and (c) show the real parts of the decoded logical density matrices at $6\,\mathrm{dB}$ and $15\,\mathrm{dB}$, respectively.
      The logical basis is ordered as $\{\ket{\bar 0 \bar 0},\ket{\bar 0 \bar 1},\ket{\bar 1 \bar 0},\ket{\bar 1 \bar 1}\}$.
      Photon recoil, photon loss, motional heating, and motional dephasing are disabled in all panels.
    }\label{noiseless-result}
    \end{figure}
    We first evaluate the intrinsic performance of the protocol in the absence of physical noise.
    Under this assumption, the remaining infidelity originates from the finite squeezing of the approximate GKP states.
    The resulting fidelity therefore provides the baseline for the present protocol with finite-energy  GKP states.
    \par
    Figure~\ref{noiseless-result}(a) shows the logical Bell-state infidelity $1-F_{\mathrm L}$ as a function of the GKP squeezing level $\Delta_{\mathrm{dB}}$.
    The infidelity decreases monotonically with increasing squeezing because the overlap between neighboring peaks of the approximate GKP states is reduced.
    At $6\,\mathrm{dB}$, the decoded logical fidelity is $F_{\mathrm L}=91.78\%$, whereas at $15\,\mathrm{dB}$ it reaches $F_{\mathrm L}=98.82\%$.
    \par
    Panels in Fig.~\ref{noiseless-result}(b) and (c) show the real parts of the decoded logical density matrices at $6\,\mathrm{dB}$ and $15\,\mathrm{dB}$, respectively.
    At $6\ \mathrm{dB}$, finite squeezing suppresses the target Bell-state coherence and produces small fractions in nontarget matrix elements.
    At $15\ \mathrm{dB}$, the central four elements approach the ideal value of $1/2$, and the remaining elements are strongly suppressed.
    These results establish the finite-squeezing baseline against which the physical noise channels are evaluated below.

    \subsection{Effect of photon loss}\label{subsec:encoding-loss}
    We next compare how photon loss affects the conditional logical states generated with the single-rail and polarization encodings introduced in Sec.~\ref{subsec:gkp_distribution}.
    To evaluate the effect of photon loss, the other noises, including photon recoil, motional heating, and motional dephasing, are not considered in this subsection.
    We denote the photon loss probability by $p_{\mathrm{loss}}=1-\eta_{\mathrm{ph}}$.
    For photon detection, we consider photodetector and accept the specified one-photon and two-photon detection patterns for the single-rail and polarization encodings, respectively.
    \par
    For the single-rail encoding, photon loss acts as amplitude damping within the vacuum--one-photon qubit space.
    The photonic state after the beam splitter contains both the one-photon state and the two-photon state.
    If one of incoming two photons is lost before the BSM, the resulting one-photon detection is indistinguishable from a desired heralding event.
    Such an event is therefore accepted as a false herald and adds an unwanted component to the conditional logical state.
    The weight of this component increases with $p_{\mathrm{loss}}$, thereby reducing the logical Bell-state fidelity.
    
    \par
    For the polarization encoding, photonic qubits lie in the one-photon subspace, and the BSM accepts the two-photon detection pattern. Namely, the photon-loss event is discarded at BSM.
    Photon loss therefore reduces the success probability but unchanges their logical state fidelity.
    
\begin{figure}[t]
      \centering
      \includegraphics[width=\columnwidth]{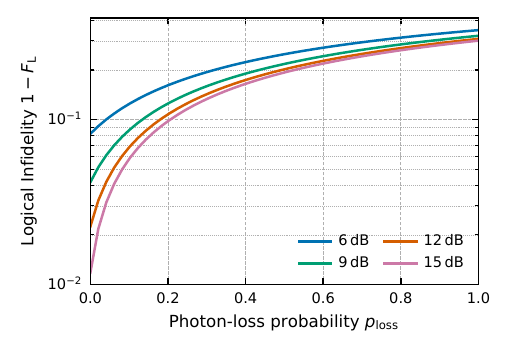}
      \caption{
          Effect of photon loss on the logical state generated with the single-rail encoding.
          The logical Bell-state infidelity $1-F_{\mathrm L}$ is shown as a function of the photon loss probability $p_{\mathrm{loss}}=1-\eta_{\mathrm{ph}}$ for GKP squeezing levels of $6$, $9$, $12$, and $15\,\mathrm{dB}$.
          The infidelity is contributed by finite squeezing and the false-herald event produced when one photon is lost from the two-photon input.
          The polarization-encoding result is not shown because, under ideal two-photon postselection, photon loss does not change the resultant state.
          Photon recoil, motional heating, and motional dephasing are disabled.
          The model assumes identical independent photon losses, photon-number-resolving detection, and no dark counts or other false-heralding processes.
      }
      \label{fig:loss-comparison}
    \end{figure}

    Figure~\ref{fig:loss-comparison} shows the logical Bell-state infidelity for the single-rail encoding as a function of $p_{\mathrm{loss}}$ at four GKP squeezing levels.
    The infidelity increases monotonically with photon loss because the contribution of false-herald events increases.
    The polarization-encoding fidelity is independent of $p_{\mathrm{loss}}$ under ideal two-photon postselection and is therefore not shown.
    We assume that the time scale of the fluctuation of polarisation rotation in two-photon-interference scheme is so slower compairing with photon transmission time sclae that we neglige from the analysis ~\cite{Krutyanskiy-EtAl2023a}.
    \subsection{Effect of photon recoil during photon generation}\label{subsec:recoil_result}
    Following the comparison above, we examine  the recoil-induced degradation of both photonic encodings.
    In this subsection, motional heating, motional dephasing, and photon loss are turned off in order to evaluate the recoil-induced degradation.
    The recoil channel is characterized by two dimensionless parameters: the Lamb-Dicke parameter $\eta$, which sets the magnitude of the momentum kick, and $\omega_{\mathrm{m}}\tau$, which determines how much the motional phase rotates during the photon emission.
    Physically, a larger $\eta$ increases the displacement amplitude in phase space, while a larger $\omega_{\mathrm{m}}\tau$ broadens the emission-time-dependent recoil direction.
    \par
\begin{figure*}[t]
    \centering
    \includegraphics[width=\textwidth]{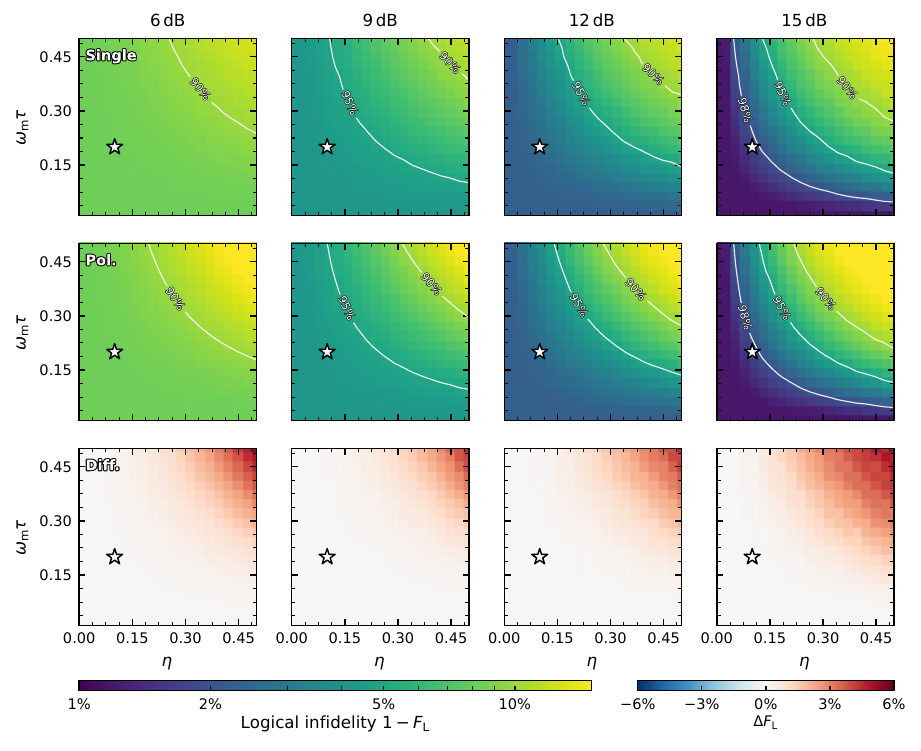}
    \caption{
      Encoding-dependent sensitivity of the remote GKP Bell state to photon recoil.
      Columns correspond to GKP squeezing levels $\Delta_{\mathrm{dB}}=6$, $9$, $12$, and $15\,\mathrm{dB}$.
      The first and second rows show the logical Bell-state infidelities $1-F_{\mathrm L}^{\mathrm{sr}}$ and $1-F_{\mathrm L}^{\mathrm{pol}}$ for the single-rail and polarization encodings, respectively.
      The horizontal and vertical axes are the Lamb--Dicke parameter $\eta$ and the dimensionless photon-wavepacket duration $\omega_{\mathrm m}\tau$.
      The first two rows use the same logarithmic color scale.
      The third row shows the difference of infidelities $\Delta F_{\mathrm L}=F_{\mathrm L}^{\mathrm{sr}}-F_{\mathrm L}^{\mathrm{pol}}$ using a diverging color scale centered at zero.
      Positive values indicate an advantage for the single-rail encoding, whereas negative values indicate an advantage for the polarization encoding.
      Photon loss, motional heating, and motional dephasing are disabled in all panels.
    }\label{fig:recoil-comparison}
    \end{figure*}

    Figure \ref{fig:recoil-comparison} shows the decoded logical Bell-state infidelity $1-F_{\mathrm{L}}$ as a function of $\eta$ and $\omega_{\mathrm{m}}\tau$ for several GKP squeezing levels.
    The infidelity increases as either $\eta$ or $\omega_{\mathrm{m}}\tau$ becomes larger.
    This behavior is consistent with the phase-space picture of the recoil channel: random emission times $t$ convert a fixed photon recoil $\hat D(i\eta)$ into a random displacement direction $\hat D\qty(-i\eta e^{i\omega_m t})$ in the rotating motional frame, thereby distorting the GKP grid.
    Unlike the Gaussian random-displacement noise associated with motional heating, the recoil channel has a fixed displacement scale set by $\eta$, while its phase is sampled from the emission-time distribution as given in Eqs.~\eqref{eq:recoil-channel-sr} and \eqref{eq:recoil-channel-dr}.
    \par
    The finite-squeezing contribution remains visible in all color maps and sets the baseline fidelity at $\eta=0$.
    Increasing the squeezing level raises this baseline fidelity, whereas increasing either $\eta$ or $\omega_{\mathrm m}\tau$ enhances the recoil-induced degradation.
    Near the experimental recoil parameters~\cite{Saha-EtAl2025} marked in Fig.~\ref{fig:recoil-comparison}, the difference of infidelities between the two encodings is small compared with their total infidelity.
    At larger $\eta$ and $\omega_{\mathrm m}\tau$, the encoding dependence becomes appreciable because recoil acts on both logical states in the polarization encoding.
    
    \par
    These results indicate that photon recoil during spontaneous emission does not by itself preclude high-fidelity remote GKP Bell-state generation.
    The recoil-induced degradation is suppressed when the Lamb-Dicke parameter is small, and the photon wavepacket duration is short compared with the motional period.
    Thus, fast spontaneous photon emission is preferable with the proposed protocol.
        
    \subsection{Effect of motional heating and dephasing during photon propagation}
    \begin{figure}[t]
    \centering
    \includegraphics[width=\linewidth]{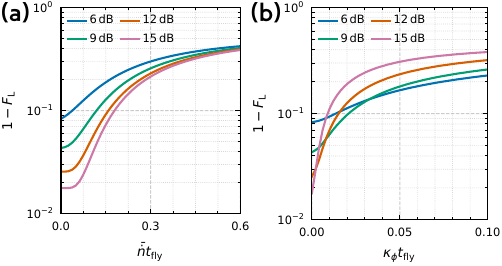}
    \caption{
        Effects of motional heating and dephasing on the decoded logical Bell state for the polarization encoding.
        Panel (a) shows the logical Bell-state infidelity $1-F_{\mathrm L}$ as a function of the dimensionless heating strength $\dot{\bar n}t_{\mathrm{fly}}$, with motional dephasing disabled.
        Panel (b) shows the infidelity as a function of the dimensionless dephasing strength $\kappa_\phi t_{\mathrm{fly}}$, with motional heating disabled.
        The curves correspond to GKP squeezing levels of $6$, $9$, $12$, and $15\,\mathrm{dB}$.
        The recoil parameters are fixed at $\eta=0.1$ and $\omega_{\mathrm m}\tau=0.2$, and photon loss is disabled.
    }\label{fig:heating_dephasing}
    \end{figure}
    We then include motional noise during the time interval $t_{\mathrm{fly}}$ between photon generation and the photonic Bell-state measurement.
    During this interval, the motional modes are affected by anomalous heating and motional dephasing. 
    Motional heating is modeled as an isotropic random-displacement channel as given in Eq.~\eqref{eq:heating-channel}, while motional dephasing corresponds to random phase-space rotations of the oscillator as given in Eq.~\eqref{eq:dephasing-channel}. 
    Both processes distort the GKP grid and can lead to logical errors after ideal GKP decoding.
    \par
    Figure \ref{fig:heating_dephasing} shows the decoded logical Bell-state infidelity as a function of the dimensionless heating and dephasing strengths, $\bar{\dot{n}}t_\text{fly}$ and $\kappa_\phi t_\text{fly}$, respectively.
    In this calculation, photon recoil is fixed to the parameters used in Fig.~\ref{fig:recoil-comparison}. 
    The fidelity decreases monotonically with both noise rates.
    Heating directly broadens the GKP peaks through random displacements, while dephasing rotates the phase-space grid in a number-dependent manner.
    As a result, both noise sources reduce the distinguishability of the logical GKP lattice points and degrade the recovered two-qubit logical state.
    \par
    The squeezing dependence differs qualitatively between the two noise channels.
    For heating, the high squeezing level suppresses the infidelity in lower heating strength, while the curves converge to the same infidelity as random-displacement noise becomes dominant.
    For dephasing, higher squeezing also lowers the infidelity but increases the sensitivity to the phase diffusion, leading to crossings between the curves.
    This enhanced dephasing sensitivity arises because larger squeezing converts the small phase rotation into large position or momentum deviation, leading to typical logical errors.
    \par
    For the representative flight time and trap parameters used in Fig. \ref{fig:heating_dephasing}, the generated remote GKP Bell state remains above $95\%$ logical fidelity when the motional heating rate is below the order of $10^{2}\ \mathrm{quanta/s}$ and the motional dephasing rate is below the order of $10\,\mathrm{Hz}$. 
    These values have already been demonstrated in current trapped-ion experiments~\cite{Matsos-EtAl2024}.
    
    \subsection{Combined noise under experimental parameters}\label{subsec:realistic-situation}
    \begin{figure*}[htbp]
    \centering
    \includegraphics[width=\linewidth]{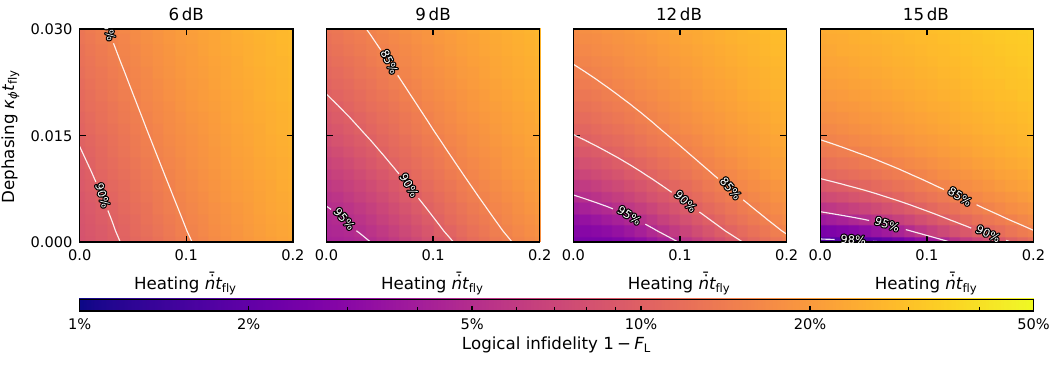}
    \caption{
      Combined effects of motional heating and dephasing on the decoded logical Bell state for the polarization encoding.
      The four panels correspond to GKP squeezing levels $\Delta_{\mathrm{dB}}=6$, $9$, $12$, and $15\,\mathrm{dB}$.
      The horizontal and vertical axes are the dimensionless heating strength $\dot{\bar n}t_{\mathrm{fly}}$ and dephasing strength $\kappa_\phi t_{\mathrm{fly}}$, respectively.
      The color scale shows the logical Bell-state infidelity $1-F_{\mathrm L}$, and the white curves are contours of constant fidelity.
      The recoil parameters are fixed at $\eta=0.1$ and $\omega_{\mathrm m}\tau=0.2$, and photon loss is not included because we assume polarisation encoding scheme.
    }\label{fig:realistic}
    \end{figure*}
    \par
    Finally, we evaluate the combined effects of photon recoil, motional heating, and motional dephasing with experimental parameters.
    We fix the recoil parameters to $\eta=0.1$ and $\omega_{\mathrm m}\tau=0.2$ and set the photon-propagation time to $t_{\mathrm{fly}}=300\ \upmu\mathrm{s}$, which corresponds to $3\ \mathrm{dB}$ attenuation by $0.14\ \mathrm{dB/km}$ fibre \cite{Tamura-EtAl2018}.
    We then calculate the decoded logical Bell-state fidelity as a function of the dimensionless heating strength $\dot{\bar n}t_{\mathrm{fly}}$ and dephasing strength $\kappa_\phi t_{\mathrm{fly}}$ for several GKP squeezing levels.
    \par
    Figure~\ref{fig:realistic} shows the combined dependence of the logical infidelity on motional heating and dephasing.
    At each squeezing level, both noise strengths increase the infidelity.
    Increasing the squeezing level reduces the finite-squeezing contribution but also makes the state more sensitive to phase diffusion as discussed in the previous subsection.
    For $t_{\mathrm{fly}} = 100\ \upmu\mathrm{s}$, $\dot{\bar n}t_{\mathrm{fly}}=0.03$ and $\kappa_\phi t_{\mathrm{fly}}=3\times10^{-3}$ at $\dot{\bar n}=300\ \mathrm{quanta/s}$ and $\kappa_\phi=30\ \mathrm{Hz}$, respectively.
    This conversion connects the dimensionless channel parameters used in the calculation to experimental noise rates.
    \section{Conclusion}\label{sec:conclusion}
    We proposed a protocol for generating remote entanglement between GKP qubits encoded in the motional modes of trapped ions.
    The internal electronic states interface the motional GKP qubits with emitted optical photons, avoiding the need for direct coherent motion--photon conversion.
    The required motional-control~\cite{Fluhmann-EtAl2019, Matsos-EtAl2024} and photonic-networking~\cite{Casabone-EtAl2013, Hucul-EtAl2015, Krutyanskiy-EtAl2023a, Liu-EtAl2026c, Moehring-EtAl2007a, Nolleke-EtAl2013, OReilly-EtAl2024a, Saha-EtAl2025, Slodicka-EtAl2013, Stephenson-EtAl2020} operations have already been demonstrated in trapped-ion experiments, making the protocol implementable by combining existing experimental techniques.
    \par
    We derived the entanglement-generation process for single-rail and polarization photonic encodings and reconstructed the resulting logical states using stabilizer subsystem decomposition.
    Finite GKP squeezing limits logical infidelity even under noiseless setting.
    The two photonic encodings have similar recoil sensitivity in the weak-recoil regime but differ in the strong-recoil regime.
    Photon loss introduces false heralds and changes the conditional logical state for the single-rail encoding, whereas for the polarization encoding, photon-loss events are rejected by the two-photon detection condition.
    We also quantified how photon recoil, motional heating, and motional dephasing modify the decoded logical Bell-state fidelity over experimental parameter ranges.
    These calculations provide a link-level characterization of the proposed remote motional-GKP entanglement protocol over experimental noise parameters.
    \par
    Several extensions remain for future work.
    In this study, we focused on GKP encoding, but other bosonic codes, including cat, squeezed-cat, and binomial codes, may exhibit different trade-offs between finite-energy error, recoil sensitivity, and experimental implementation~\cite{Cochrane-Milburn-Munro1999,Michael-EtAl2016,Omkar-Teo-Jeong2020,Lee-EtAl2024a,Kiryu-EtAl2025}.
    We modeled photon generation using spontaneous emission, while a cavity-assisted implementation could modify the photon-collection efficiency and recoil dynamics~\cite{Tanji-EtAl2024,Kikura-EtAl2025b}.
    Our evaluation used logical density matrices reconstructed by SSSD and therefore corresponds to ideal GKP decoding.
    An experimental implementation of GKP error correction requires repeated syndrome-extraction operations with finite control errors and time overhead~\cite{deNeeve-EtAl2022}.
    Including these imperfections in the complete entanglement-generation sequence is an important direction for future work.
    
    \section*{Data Availability}
    The data that support the findings of this study are available from the corresponding author upon reasonable request.
    
    \begin{acknowledgments}
    This work is supported by JST Moonshot R\&D (Grant Nos. JPMJMS256E, JPMJMS256G), JST ASPIRE (Grant No.: JPM-JAP2427), and Keio University Program for the Advancement of Next Generation Research Projects (Type A), JSPS KAKENHI (Grant No. JP25K01292), MEXT Q-LEAP (Grant No. JPMXS0118067395), JST CREST (Grant No. JPMJCR24A5).
    K.T. and S.K. are supported by JST SPRING (Grant Number JPMJSP2123).
    \end{acknowledgments}
    
    
  \bibliography{ref}
\end{document}